# Multiple-pseudogap phases in hydrogen-doped LaFeAsO system

A. Nakamura[1], T. Shimojima[1], T. Sonobe[1], K. Ishizaka[1], W. Malaeb[2], S. Shin[2], S. Iimura[3], S. Matsuishi[4], H. Hosono[3,4]

[1]*Quantum-Phase Electronics Center and Department of Applied Physics,*
*The University of Tokyo, Hongo, Tokyo 113-8656, Japan*
[2]*Institute of Solid State Physics, The University of Tokyo, Kashiwa, Chiba 227-8581, Japan*
[3]*Laboratory for Materials and Structures, Tokyo Institute of Technology, Yokohama 226-8503, Japan*
[4]*Materials Research Centere for Element Strategy, Tokyo Institute of Technology, Yokohama 226-8503, Japan*

The low energy electronic structure of LaFeAsO$_{1-x}$H$_x$ ($0.0 \leq x \leq 0.60$), the system which exhibits two superconducting domes in its phase diagram, is investigated by utilizing the laser photoemission spectroscopy. From the precise temperature-dependent measurement of the spectra near the Fermi level, we find the suppression of the density of states with cooling, namely the pseudogap formation, for all doping range. The pseudogap in the low $x$ range (i.e. the first superconducting dome regime) gets suppressed with increasing $x$, more or less similarly to the previous results in F-doped LaFeAsO system. On the other hand, the pseudogap behavior in the second superconducting dome regime at high-$x$ becomes stronger with increasing the H-doping level. The systematic doping dependence shows that the pseudogap is enhanced toward the both ends of the phase diagram where the different types of antiferromagnetic order exist.

## I. INTRODUCTION

Superconductivity in iron arsenides was discovered by applying F-doping to antiferromagnetic LaFeAsO, with a maximum superconducting critical temperature ($T_c$) of 26 K. [1] The $Ln$FeAsO ($Ln$ = lanthanide) system tends to show high $T_c$ as compared to other iron pnictide systems. Specifically, SmFeAsO$_{1-x}$F$_x$ displays one of the highest $T_c$ among the iron pnictides [2]; 55 K at $x \sim 0.10$. Recent developments in the hydrogen substitution method [3, 4, 5] have greatly increased the electron-doping limit of the $Ln$FeAsO system from $x \sim 0.2$ up to $x \sim 0.6$. [6] In LaFeAsO$_{1-x}$H$_x$, the double-dome shaped superconducting phase appears as a function of $x$ as shown in Fig. 1; the low-$x$ superconducting dome (SC1) at $x$ = 0.05 - 0.20 with maximum $T_c$ of 29 K, and the high-$x$ superconducting dome (SC2) at $x$ = 0.20 - 0.42 with maximum $T_c$ of 36 K. These two superconducting domes seem to merge into one by applying high pressure [7] or by substituting the La ion by other lanthanides of smaller ion radius, such as Ce, Sm, and Gd [6]. In these cases, the maximum $T_c$ becomes highly enhanced as compared to the LaFeAsO$_{1-x}$H$_x$ system in ambient pressure. Similar two-dome superconducting phase diagrams are also obtained in LaFe(As$_{1-x}$P$_x$)O$_{1-y}$F$_y$ [8] and SmFeAs$_{1-y}$P$_y$O$_{1-x}$H$_x$ [9], thus indicating the possible competition or cooperation of two different superconducting mechanisms that are inherent in $Ln$FeAsO system [10, 11]. From this viewpoint, the investigation of SC1 and SC2 in LaFeAsO$_{1-x}$H$_x$ is important for revealing the mechanism that leads to high $T_c$ value in the $Ln$FeAsO family.

The phase diagram of LaFeAsO$_{1-x}$H$_x$ is further characterized by two antiferromagnetic (AF) states, namely the AF1 ($x < 0.05$) and AF2 ($0.4 < x$), as shown in Fig. 1. AF2 has been clarified by nuclear magnetic resonance (NMR) [12, 13], inelastic neutron scattering [14], and muon spin rotation ($\mu$SR) measurements.[15] The experiments show that AF1 and AF2 exhibit different magnetic ordering vectors, magnetic moments, and antiferromagnetic transition temperatures ($T_N$). It is also worth noting that there are two types of structural transitions on cooling below the structural phase transition temperature ($T_S$) from tetragonal to orthorhombic, whose doping dependence is similar to that of $T_N$. Such a rich phase diagram implies a possible variety of the normal-state electronic properties as the background for SC1 and SC2 domes. According to the density functional theory calculations [6, 16, 17, 18, 19, 20], LaFeAsO$_{1-x}$H$_x$ system has hole and electron Fermi surfaces at the center and the corners of the Brillouin zone, respectively, which are composed of multiple Fe $3d$ orbitals. Upon electron doping from $x = 0.0$ to $x = 0.40$, the shape of the Fermi surface drastically changes. Consequently, the spin fluctuations connecting between the Fermi surfaces of $YZ$/$ZX$ orbitals are expected to develop in the SC1 region while those between the Fermi surfaces of $X^2 - Y^2$ orbitals should become dominant in SC2 [19, 21] ($X$/$Y$ and $Z$ correspond to the tetragonal axes $a^T$ and $c^T$). Theoretical studies considering both spin and orbital susceptibilities, on the other hand, have proposed the simultaneous evolution of the spin and orbital fluctuations for both SC1 and SC2 phases. [18]

Experimentally, photoemission spectroscopy studies on F-doped LaFeAsO$_{1-x}$F$_x$ ($0 \leq x \leq 0.14$) were employed soon after its discovery, to investigate the normal-state electronic structure. [22, 23, 24, 25] Some of them reported the pseudogap evolving from the temperature above the SC1 phase transition, which was attributed to the precursor of the antiferromagnetic gap or the antiferromagnetic spin fluctuations. [23] However, there has been no report on LaFeAsO$_{1-x}$H$_x$ system until now, where high-$x$ SC2 dome is available. Recently, an NMR measurement revealed that the spin relaxation rate $1/T_1T$ gets strongly suppressed with cooling below $T^*$ ($T^* > T_N$) in the heavily doped regime ($x > 0.4$). [26] Such a deviation from the Curie-Weiss behavior may suggest the possible emergence of a pseudogap also in SC2 region. Considering that the antiferromagnetic ordered states at $x \sim 0.0$ and $x \sim 0.50$ show different magnetic and structural properties [15], the systematic investigation of the electronic structure in a wide range of the phase diagram will provide information on the origin of the two-dome superconductivity and its possible relation to the pseudogaps.

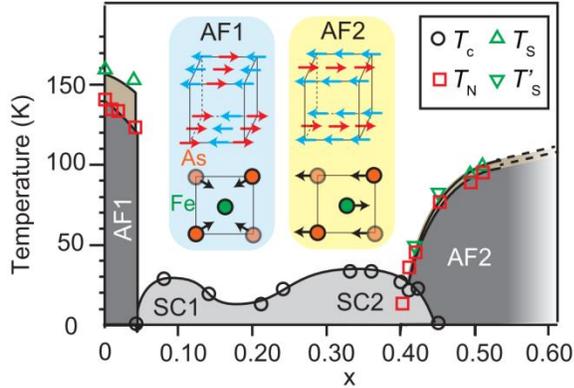

FIG. 1. Phase diagram of LaFeAsO$_{1-x}$H$_x$. [6, 12, 13, 15] $T_S'$ indicates the temperature where the $c$ axis length shows an upturn as a function of temperature, as observed by X-ray diffraction measurements [15]. The magnetic structure and the displacements of the Fe and As atoms in the AF1 and AF2 ordered phases are also shown in inset [15].

In this study, we use angle-integrated photoemission spectroscopy (AIPES) to investigate the polycrystalline LaFeAsO$_{1-x}$H$_x$ in a wide range of temperatures (6 - 300 K) and compositions (0.0 ≤ $x$ ≤ 0.60). The temperature-dependent density of states near the Fermi energy ($E_F$) can be precisely acquired by using the laser-AIPES. We find a new pseudogap in the high-$x$ region ($x$ = 0.35 - 0.60), evolving from the temperatures above SC2 and AF2 phase transitions. The $x$-dependence of the pseudogap in SC2-AF2 region is clearly distinguished from the pseudogaps in low-$x$ AF1-SC1 region. The V-shaped doping dependence of the pseudogap temperature ($T_{PG}$) in the whole phase diagram indicates that these pseudogaps originate from respective electronic ground states of the AF1 ($x \sim 0.0$) and AF2 ($x \sim 0.50$), respectively, which may be also crucial for the occurrence of two-dome superconductivity.

## II. METHODS

Polycrystalline LaFeAsO$_{1-x}$H$_x$ ($x$ = 0.0, 0.10, 0.20, 0.35, 0.50, 0.60) samples were synthesized as described in Ref. 6.

High-energy-resolution laser AIPES measurements were performed with a spectrometer built using a VG-Scienta R4000 electron analyzer and an ultraviolet-laser of 6.994 eV as a photon source at the Institute for Solid State Physics, University of Tokyo. [27, 28] The energy resolution was set to about 6 meV to obtain a high count rate of photoelectrons. Fermi energy of the samples was referenced to that of a gold film evaporated onto the sample holder. All the polycrystalline samples were fractured *in situ* at 200 K in an ultra-high vacuum better than $1 \times 10^{-10}$ Torr. We confirmed the reproducibility of the temperature-dependent AIPES spectrum by measuring it during the temperature cycle.

## III. RESULTS AND DISCUSSIONS

Figure 2 shows the photoemission spectra near the Fermi energy at $x$ = 0.0 (AF1), 0.10 (SC1), 0.35 (SC2), and 0.50 (AF2). The raw spectra shown in Figs. 2(a)–2(d) were normalized by the spectral intensity integrated between the binding energies of 70 meV and 100 meV. All the spectra show an almost linear slope toward the higher binding energy, being consistent with previous photoemission measurements on F-doped LaFeAsO$_{1-x}$F$_x$. [22, 23, 24, 25] Apparently, the spectral intensity at the Fermi energy decreases with cooling for all compositions. To remove the contributions of the Fermi–Dirac distribution and focus on the temperature dependence of the density of states itself, the spectra were symmetrized with respect to the Fermi energy, as shown in Figs. 2(i)–2(l). The thin blue curves overlaid on respective red curves are those obtained at the lowest temperature i.e., 12 K for $x$ = 0.0 (AF1), 7 K for $x$ = 0.10 (SC1), 8 K for $x$ = 0.35 (SC2), and 10 K for $x$ = 0.50 (AF2). The black triangles indicate the energy position where the thin blue curve deviates from the red curve. Here, we can distinguish a suppression of the density of states near the Fermi energy with cooling, for all the samples. We also note that the temperature dependence of the symmetrized spectra is qualitatively similar to that of the raw data divided by the Fermi-Dirac function, shown in Figs. 2(e)–2(h). It indicates that the process of symmetrization does not affect the temperature dependent spectral suppression.

When we look at the energy positions of the triangle markers as a function of temperature, we notice two different behaviors. At high temperature, the black triangles show a monotonic decrease of the energy scale with cooling. In contrast, below a certain temperature [here named $T_0$, see Fig. 2(i)-2(l) for their values], the positions of the black triangle markers become nearly temperature-independent. The spectral suppression with the temperature-dependent energy scale obtained at $T > T_0$ was also reported in a previous photoemission study on LaFeAsO$_{1-x}$F$_x$. [22] Since the energy scale is nearly proportional to $k_BT$ ($k_B$ : Boltzman constant), this observation was discussed in terms of the thermal effect in

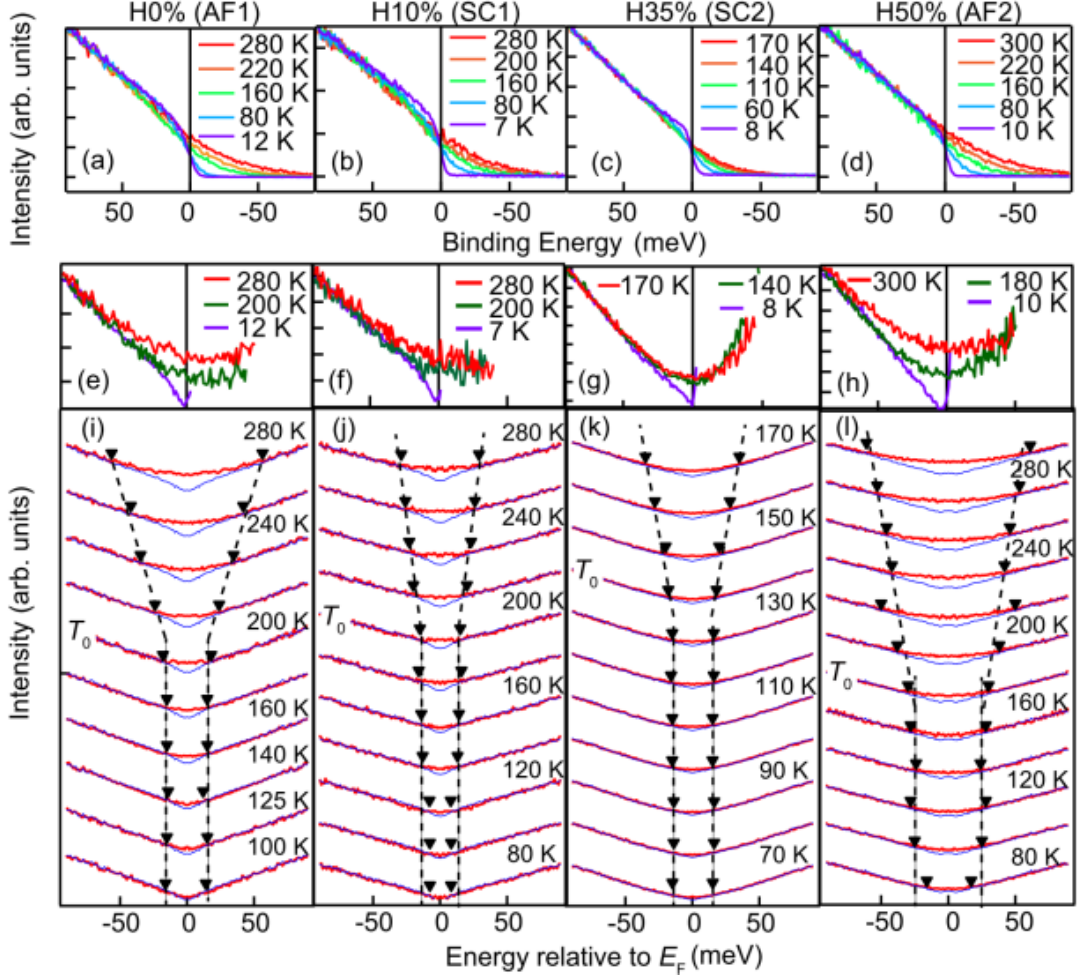

FIG. 2. (a)–(d) Temperature dependence of AIPES raw spectra near the Fermi energy for $x = 0.0$ (AF1), 0.10 (SC1), 0.35 (SC2), and 0.50 (AF2), respectively. (e)–(h) Temperature dependence of the AIPES spectra divided by the Fermi-Dirac distribution function for $x = 0.0$, 0.10, 0.35, and 0.50, respectively. (i)–(l) The symmetrized AIPES spectra for $x = 0.0$ (AF1), 0.10 (SC1), 0.35 (SC2), and 0.50 (AF2), respectively. The thick red curves show the data at respective temperatures, whereas the thin blue curves denote those at the lowest temperatures; 12 K for $x = 0.0$, 7 K for $x = 0.10$, 8 K for $x = 0.35$, and 10 K for $x = 0.50$, respectively. The black triangles indicate the energy where the thick red curves deviates from the thin blue curves.

the semi-metal-like electronic structure of the compound. [22] On the other hand, the spectral suppression with the temperature-independent energy scale occurring at $T < T_0$ is similar to the pseudogap formation in cuprates and pnictides. [29, 30]

Here we focus on the low temperature region ($T < T_0$) where the $k_B T$-dependent features are not dominant and can be more or less excluded. The temperature dependence of the raw spectra below $T_0$ is shown in Figs. 3(a)–3(d). The insets show the magnified spectra at the Fermi energy, whose intensities decrease on cooling. The symmetrized spectra were further normalized by those at $T_0$, and are shown in Figs. 3(e)–3(h). The magnifications of these spectra at several temperatures are displayed in Figs. 3(i)–3(l). For $x = 0$ [Fig. 3(e)], the spectra show a slightly gapped area gradually appearing at 160–155 K on cooling, in the energy range up to ~35 meV. We also find that the integrated area from the Fermi energy to the binding energy of 100 meV is not conserved when the temperature is varied. This suggests that the decreased spectral weight is redistributed over a wide energy range, similar to the case of Kondo insulators. [31] The estimated energy scale of the gap-like structure, displayed in Fig. 3(i), is indicated by the broken lines in Fig. 3(e). This energy scale is temperature-independent, being distinctly different from the $k_B T$-dependent feature appearing at higher temperature. Considering that the parent LaFeAsO ($x = 0$) is known to exhibit an antiferromagnetic transition at 140 K, the gap-like depression in the density of states evolving already at 160 K should not correspond exactly to the antiferromagnetic gap itself. We identify this gap-like feature as a pseudogap (PG1), and estimate its characteristic temperature ($T_{PG1}$) to be 160 ± 5 K. Regarding the AF transition at 140 K, we could not separately find any additional feature representing the antiferromagnetic gap, which might be because of the comparable energy scales of the pseudogap ($\Delta_{PG}$) and antiferromagnetic gaps.

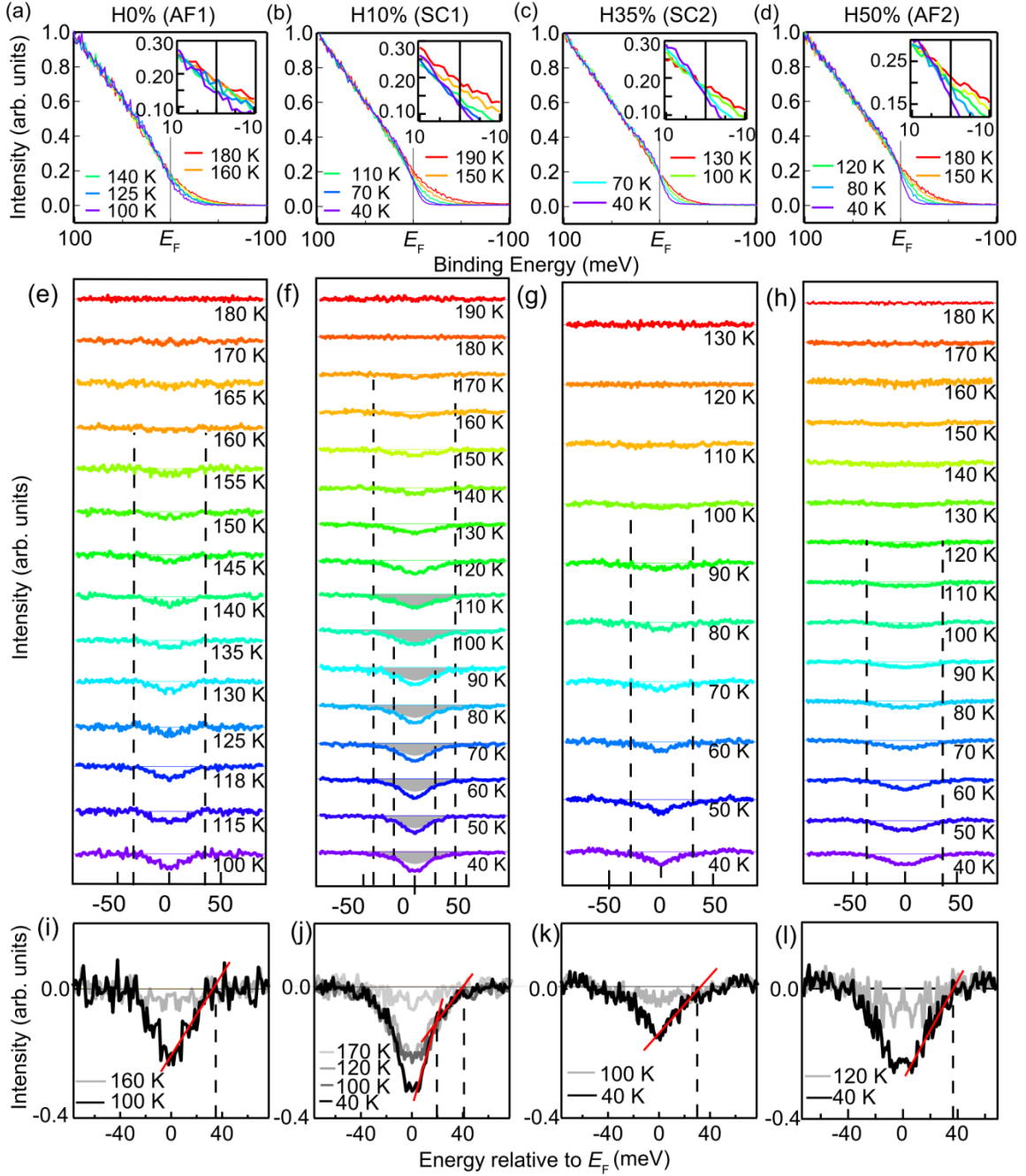

FIG. 3. (a)–(d): Temperature dependence of raw AIPES spectra for $x = 0.0$, 0.10, 0.35, and 0.60, respectively. (e)–(h): Normalized spectra for $x = 0.0$, 0.10, 0.35 and 0.60, respectively. The broken lines indicate the energy scale of the pseudogap, $\Delta_{PG}$. The shaded area in (f) represents the gapped area at 120 K, used to highlight the smaller pseudogap formation below 120 K. (i)-(l) The magnification of the normalized spectra at some specific temperatures are shown for $x = 0.0$, 0.10, 0.35, and 0.50, respectively. The slope of the gapped feature was fitted by a linear function, and its intersection with the zero level of the intensity was defined as the value of $\Delta_{PG}$ (broken line). At $x = 0.10$, the spectrum at 40 K shows a deviation from that at 120 K within the energy scale of 20 meV, representing the formation of the smaller additional pseudogap, PG1L.

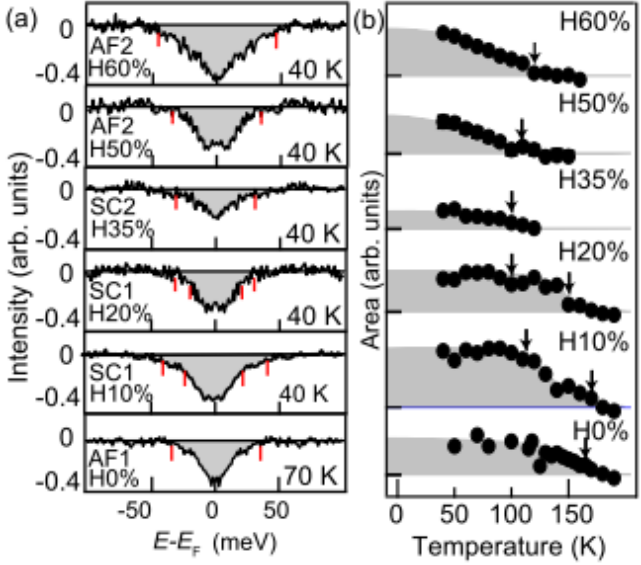

FIG. 4. (a) Normalized AIPES spectra for all x, recorded at low temperature (40 or 70 K), are plotted. The gray shade show the gapped area which represents the depression of the spectral intensity due to the pseudogap formation. The red vertical bars show the energy scales of the pseudogaps, $\Delta_{PG1L}$, $\Delta_{PG1H}$, and $\Delta_{PG2}$ as estimated from Figs. 3(e)-3(h). (b) Temperature dependence of the gapped area, determined by integrating the gray shaded region in (a). The black arrows indicate $T_{PG}$, as estimated from Figs. 3(e)–3(h).

In the case of $x = 0.10$, the optimal composition of SC1, we find the evolution of the pseudogap below ~170 K with energy scale of $\Delta_{PG1}$ ~40 meV, as shown in Fig. 3(f). In addition, another smaller gap of 20 meV appears below ~100 K. This feature is discernible in Fig. 3(f), by comparing the temperature dependent spectra to the gray shaded area depicting the depression of the spectra at 120 K. This suggests the existence of two pseudogaps with lower $T_{PG1L}$ (PG1L) and higher $T_{PG1H}$ (PG1H) in LaFeAsO$_{1-x}$H$_x$. Similar PG features with two energy scales are also observed for $x = 0.20$, with $T_{PG1H} = 150$ K, $\Delta_{PG1H}$ ~ 30 meV and $T_{PG1L} = 90$ K, $\Delta_{PG1L}$ ~ 20 meV (not shown). Such two energy-scaled pseudogaps had also been reported in a previous AIPES study on LaFeAsO$_{1-x}$F$_x$, [25] ($x < 0.10$), thus indicating that they indeed represent the normal state of the SC1 region, regardless of the dopants. The possible origins of these pseudogaps will be discussed later.

For the SC2 and AF2 regions ($x > 0.35$), we also find the pseudogap evolution on cooling (PG2). As shown in Figs. 3(g) and 3(k), the PG2 with $\Delta_{PG2} = 30$ meV starts to evolve at $T_{PG2} = 100$ K for $x = 0.35$ (SC2). Partly because the PG2 is smaller compared to $x = 0.10$, it is difficult to discuss the possible presence of an additional low-temperature pseudogap in the present data. The observed pseudogap feature seems to be enhanced with increasing $x$ to $x = 0.5$, as shown in Figs. 3(h) and 3(l). The energy and temperature scales of the pseudogap are $\Delta_{PG2} = 35$ meV and $T_{PG2} = 120$ K for $x = 0.50$, and $\Delta_{PG2} = 45$ meV and $T_{PG2} = 130$ K for $x = 0.60$ (not shown). According to the $\mu$SR measurements [15], $T_N$ at $x = 0.50$ (AF2) was estimated to be approximately 90 K. The observed $T_{PG2}$ is again higher than $T_N$, which indicates that the gap features observed at $x = 0.50$ and 0.60 are not directly associated with the antiferromagnetic gap itself.

Now, we discuss the doping dependence of the pseudogaps for $0.0 \leq x \leq 0.60$, which includes the AF1, SC1, SC2, and AF2 phases. Figure 4(a) shows the normalized spectra at 40–70 K for each doping, obtained similarly with those in Figs. 3(e)–3(h). The red markers indicate the energy scales of the pseudogaps. We further estimated the temperature-dependent gapped area by integrating the gray shaded area in Fig. 4(a), and the result is shown in Fig. 4(b). The arrows in Fig. 4(b) indicate $T_{PG}$, which were estimated in Figs. 3(e)–3(h). For PG1 in the SC1 region ($x = 0.10, 0.20$), there are two pseudogap temperatures, $T_{PG1H}$ and $T_{PG1L}$, as discussed above. As we can see in Fig. 4(b), both $T_{PG1H}$ and $T_{PG1L}$ decrease with doping, similar to the findings of previous studies on LaFeAsO$_{1-x}$F$_x$. [23] With further doping, the onset temperature of the pseudogap decreases to 100 K at $x = 0.35$ (SC2). However, it increases again to 130 K at $x = 0.60$ (AF2). Thus, PG2 for the SC2-AF2 region clearly shows a different doping-dependent behavior compared to PG1. The evolution of PG2 for the SC2-AF2 region can be also recognized in $\Delta_{PG2}$ and the magnitude of the gapped area, both of which monotonically increase with increasing $x$ from 0.35 to 0.60. From this doping dependence, we can conclude that PG1 and PG2 exist throughout the phase diagram of LaFeAsO$_{1-x}$H$_x$, which seem to be enhanced toward AF1 and AF2, respectively.

Here we confirm the bulk superconductivity of LaFeAsO$_{1-x}$H$_x$ by AIPES measurements below $T_c$. Figures 5(a) and 5(b) show the AIPES spectra below and above $T_c$ for $x = 0.10$ (SC1) and $x = 0.35$ (SC2), respectively. We can see a slight decrease in the spectral weight near Fermi energy, which is further emphasized in the symmetrized spectra shown in Figs. 5(c) and 5(d). By dividing the spectra in the superconducting state by that in the normal state, the superconducting gap feature is extracted, as shown in Figs. 5(e) and 5(f). The observation of the superconducting gap opening below the superconducting transition temperature confirms that the laser-AIPES spectra represent the bulk superconducting state. The observed sizes of the superconducting gap for optimally doped compositions of SC1 and SC2 were estimated to be ~5 meV. Within the experimental error, the difference in the superconducting gap size between SC1 and SC2 was not clarified. The energy scale of the superconducting gap observed in our measurements is consistent with the size of the largest multiple superconducting gap reported in various experiments, such as the point-contact Andreev reflection [32], scanning tunneling spectroscopy [33],

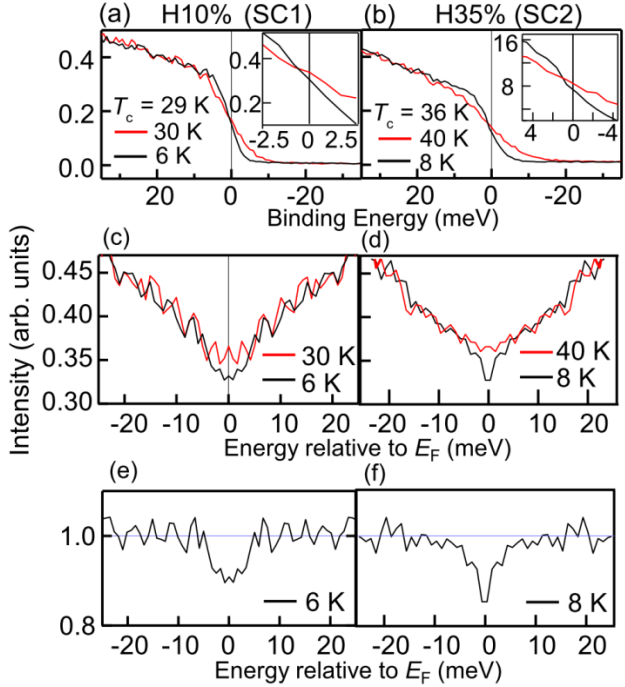

FIG. 5. Photoemission spectra below and above $T_c$ for $x = 0.10$ (a) and $x = 0.35$ (b). Symmetrized spectra for $x = 0.10$ (c) and $x = 0.35$ (d). Symmetrized spectra further normalized by that above $T_c$ for $x = 0.10$ (e) and $x = 0.35$ (f).

nuclear quadrupole resonance [34], and far-infrared reflectivity [35] experiments. We note that the magnitude of the superconducting gap is at least 3 times smaller than that of the pseudogap, indicating that the pseudogaps do not reflect the pairing precursor, as suggested in some cuprates. [36, 37]

Based on our result, all $T_{PG}$ values are summarized in the phase diagram of Fig. 6 with the orange bars. For comparison, the $T_{PG}$ values for F-doped samples are also displayed by the open squares and open triangles, taken from Ref. 23 and Ref. 25, respectively. As mentioned above, both $T_{PG1H}$ and $T_{PG1L}$ are suppressed with H-doping. The observation of the two pseudogap temperatures in SC1 region is consistent with the previous AIPES study of LaFeAsO$_{1-x}$F$_x$ [25], whereas $T_{PG1H}$ obtained in the present work may be slightly higher than that reported for LaFeAsO$_{1-x}$F$_x$. Reference 10 reported a monotonic decrease of $T_{PG1L}$ on increasing x from 0.0 to 0.14, which seems to be smoothly connected to $T_N$. Such behavior is also consistent with our result for LaFeAsO$_{1-x}$H$_x$, which may suggest that PG1L is commonly related to the AF1 phase. In addition to PG1H and PG1L, we observed PG2 in the high-$x$ SC2 and AF2 regions, which seems to be enhanced with doping H. This suggests that PG2 originates not from AF1, but from another electronic ground state at the higher doping region. The present phase diagram clarifies the existence of PG1(L,H) and PG2, which develop toward both ends of the diagram; i.e., the AF1 and AF2 ordered states around $x \sim 0.0$ and 0.5.

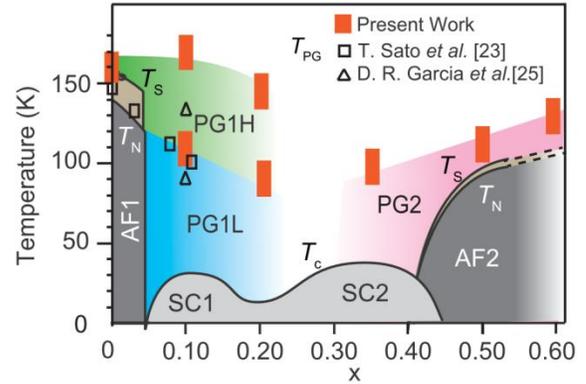

FIG. 6. Phase diagram showing the pseudogaps of LaFeAsO$_{1-x}$H$_x$. The orange bars represent the $T_{PG}$ values of H-doped LaFeAsO, as estimated from the present laser-AIPES measurements. The open squares and triangles represent $T_{PG}$ values estimated by previous AIPES studies on F-doped LaFeAsO$_{1-x}$F$_x$. [23, 25] The $T_c$, $T_N$, and $T_S$ values were taken from Ref. 6 and Ref. 15.

The pseudogaps in iron pnictides have been experimentally and theoretically discussed in relation to the spin/orbital fluctuations. Considering the smooth doping dependence of $T_{PG1L}$ that follows after $T_N$ in AF1 phase, the spin fluctuation derived from hole and electron Fermi surface nesting is a possible candidate for the origin of PG1L, as also raised in the previous photoemission studies. [23, 25] PG1H, on the other hand, had been discussed in association with the structural phase transition in F-doped LaFeAsO [25]. For the present investigation of LaFeAsO$_{1-x}$H$_x$, $T_{PG1H}$ seems to be higher than the structural transition temperature in the AF1 region. If this is the case, an electronic nematicity evolving around 175 K detected by in-plane resistivity [38] may be also playing a crucial role for pseudogap formation, as mentioned in Ref. 30. Regarding PG2, on the other hand, the enhancement of $T_{PG2}$ toward high-x implies that PG2 is related to the AF2 phase. Such pseudogap formation may correspond to the evolution of the spin/orbital fluctuations peculiar to SC2 region, as suggested by theoretical studies. [6, 16, 17, 18, 19, 20, 21] Actually, spin fluctuations with different wave numbers at $x = 0.0$ and $x = 0.40$ have been detected by inelastic neutron scattering measurements [14], while both type of fluctuations are suppressed in the middle ($x \sim 0.20$) region. This is qualitatively consistent with the crossover from PG1 to PG2 as we observed. An NMR measurement, on the other hand, proposed that the the orbital degrees of freedom or the orbital ordering may be in charge of the pseudogap behavior in the SC2 region [26]. To more solidly clarify the PG1 and PG2 states, and discuss how they overlap or crossover to each other, the precise electronic structures

## IV. CONCLUSIONS

In summary, we performed laser AIPES on LaFeAsO$_{1-x}$H$_x$ ($0 \leq x \leq 0.60$) for a wide temperature region, and observed peculiar pseudogaps of energy scales 20 ~ 45 meV existing throughout the phase diagram. While the pseudogap in $x \leq 0.20$ (PG1) becomes suppressed on doping away from x = 0, PG2 in $x \geq 0.35$ enhances toward the high-x region. It thus indicates the different origins of PG1 and PG2, possibly related to two antiferromagnetic phases, AF1 (x = 0) and AF2 (x = 0.6). We further observed the superconducting gap at SC1 and SC2 with an energy scale much smaller than $\Delta_{PG}$, indicating that pseudogap formation is not due to precursor pairing. The present result indicates that there are two types of spin/orbital fluctuations existing in LaFeAsO$_{1-x}$H$_x$ system, which should lead to the two pseudogap phases and possibly also to the two superconducting domes.

## ACKNOWLEDGEMENTS

We thank S. Ideta for valuable discussions. A. Nakamura acknowledges support by Advanced Leading Graduate Course for Photon Science (ALPS) at the University of Tokyo. This research was partly supported by the Photon Frontier Network Program of the MEXT; Research Hub for Advanced Nano Characterization, The University of Tokyo, supported by MEXT, Japan; and Grant-in-Aid for Scientific Research from JSPS, Japan (KAKENHI 15H03683 and 16K133815).

and dynamical magnetic properties using single crystals remain to be investigated in future.